\documentclass[twocolumn,showpacs,preprintnumbers,amsmath,amssymb,prb,superscriptaddress]{revtex4}
\usepackage{graphicx}
\begin{document}

\preprint{Submitted to Physical Review B}

\title{Memory effects in ac hopping conductance in the
  quantum Hall effect regime: Possible manifestation of DX$^-$
  centers}
\author{I. L. Drichko}
\email{Irina.L.Drichko@pop.ioffe.rssi.ru}
\author{A. M. Diakonov}
\author{I. Yu. Smirnov}
\affiliation{A. F. Ioffe  Physico-Technical Institute of Russian
  Academy of Sciences, 194021
  St. Petersburg, Russia}


\author{V. V. Preobrazenskii}
\author{A.I. Toropov}
\affiliation{Institute of Semiconductor Physics, Siberian division of Russian
  Academy of Sciences, Novosibirsk, Russia}



\author{Y. M. Galperin}
\email{iouri.galperine@fys.uio.no}
\affiliation{Department of Physics, University of Oslo, PO Box 1048
  Blindern, 0316 Oslo, Norway}
\affiliation{A. F. Ioffe  Physico-Technical Institute
of Russian
  Academy of Sciences, 194021
  St. Petersburg, Russia}
\date{\today}
\begin{abstract}
Using simultaneous measurements of the attenuation and velocity of
surface acoustic waves propagating along
GaAs/Al$_{0.3}$Ga$_{0.7}$As heterostructures, complex ac
conductance of the latters has been determined. In the magnetic
fields corresponding to the middles of the Hall plateaus both the
ac conductance, $\sigma (\omega)$, and the sheet electron density,
$n_s$,  in the two-dimensional conducting
layer turn out to be dependent on the samples' cooling
rate. As a result, the sample ``remembers'' the cooling conditions. 
The complex conductance is  strongly dependent on an infrared
illumination which also changes both $\sigma (\omega)$ and
$n_s$. Remarkably, the correlation between  $\sigma (\omega)$ and
$n_s$ is \emph{universal}, i.~e. it is independent of the way to
change these quantities.  
The results are
attributed to two-electron defects (so-called $DX^-$ centers)
located in the Si doped layer.
\end{abstract}
\pacs{72.20.-i; 72.30.+q; 72.50.+b; 73.43.-f; 73.63.-b}
\maketitle
\section{Introduction} \label{Introduction}

As well known,~\cite{book}  in the quantum Hall effect (QHE)
regime magnetic field dependences of off-diagonal, $\sigma_{xy}$,
and diagonal, $\sigma_{xx}$, components of dc conductivity tensor
are very much different. Namely, $\sigma_{xy}$ shows a set of flat
plateaus with abrupt steps taking place at half-integer values of
the filling factor $\nu=2\pi n_s a_H^2$ where $n_s$ is sheet
density of two-dimensional electron gas (2DEG) while
$a_H=(c\hbar/eH)^{1/2}$ is the quantum magnetic length. On the
contrary, $\sigma_{xx}^{(dc)}$ is extremely small at the plateaus
and has sharp maxima at the steps between the Hall
plateaus, i. e. at \emph{half-integer} filling
factors. The conventional explanation is that at a half-integer
$\nu$  electronic states the Fermi level are
\emph{extended} while apart of the half-integer values of $\nu$
they are \emph{localized}.

A powerful way to investigate the interplay between extended and
localized states is an analysis of \emph{ac} conductivity by 
acoustical methods.~\cite{gen_ref} A surface acoustic wave (SAW)
propagating in a vicinity of a 2DEG layer produces a wave of
electric field penetrating the 2DEG. This wave creates currents
which, in turn, produce a feedback to the SAW. As a result, both
SAW attenuation $\Gamma$ and velocity $V$ depend on the properties
of the 2DEG. In particular, simultaneous measurements of
attenuation and velocity of SAW provide a unique possibility to
determine \emph{complex} ac conductivity,
$\sigma_{xx}(\omega)=\sigma_1(\omega)-i \sigma_2(\omega)$, as a
function of external magnetic field $H$ and SAW frequency $\omega$.
Furthermore, the magnetic field dependence of complex conductivity
provides an information both on the extended and localized states, as
well as on metal-to-insulator transition.

As we observed earlier,~\cite{1,2} near the steps on Hall
conductance, i. e. at half-integer $\nu$, imaginary part of the
complex ac conductance is small while 
its real part coincides with the dc transverse conductance,
$\sigma_{xx}^{(dc)}$. However, in the magnetic fields
corresponding to regions near the middles of the Hall plateaus,
i.~e. at small integer $\nu$, the difference between
$\sigma_{xx}(\omega)$ and $\sigma_{xx}^{(dc)}$ turns out to be
crucial. Namely, $\sigma_{xx}^{(dc)}$ is extremely small while
both $\sigma_1 (\omega)$ and $\sigma_2 (\omega)$ are measurable
quantities and $\sigma_2 (\omega) \gg  \sigma_1 (\omega)$.
Furthermore, up to our experimental accuracy, the dissipative
conductivity, $\sigma_1(\omega)$ is proportional to SAW frequency
and weakly dependent on the temperature. According to
Ref.~\onlinecite{3}, these facts lead to the conclusion that the
mechanism of ac conductance is \emph{hopping}.

In our experiment, the SAW is induced by inter-digital transducers at
the surface of a piezoelectric LiNbO$_3$ plate. Samples are layered structures
placed on  the plate, so the interface layer is located at some
distance from the piezoelectric surface.
 If the conductance of the interface layer is low, then the ac electric
field produced by the SAW decays as $\sim e^{-k |z|}$ where
$k=\omega/V$ is the SAW wave vector while $z$ is the distance from
the propagation surface. In the following we assume that both the
piezoelectric plate and the samples' layers are parallel to the
$xy$-plane and magnetic field ${\bf H}$ is directed along
$z$-axis. Consequently, it is the region  of the thickness
$\approx k^{-1}$ that contributes to acoustic properties of the
system. However, a perfectly conducting layer placed at a distance
$ \ll k^{-1}$ above the surface \emph{screens} the ac electric
field and confines the electric current inside the layer. As a
result, in this case both SAW attenuation and velocity are
determined mostly by the processes inside the interface layer. An
estimate for the dimensionless parameter discriminating between
the cases of weak and strong screening is the ratio
$|\sigma_{xx}|/V$. Since $\sigma_{xx}$ is a strong function of
magnetic field, variation of the magnetic field changes the screening of
the ac electric field. As a result, at different magnetic fields
different layers of the sample provide the dominant contribution
to $\sigma_{xx}(\omega)$. In particular, at the Hall
plateaus ``shunting'' of the interface layer by a doped one can be dominant.
 A more detailed analysis~\cite{4} for Si $\delta$-doped
GaAs/Al$_{0.3}$Ga$_{0.7}$As heterostructures  has shown that it is
the case. Namely, at the Hall plateaus the doped layer's
contribution to $\sigma_1 (\omega)$ turns out to be important. At
the same time, its contribution to the dc transverse conductance
remains negligibly small.

A procedure outlined in Ref.~\onlinecite{4} allowed us to separate
the contribution in the ac conductance in QHE regime of the
interface layer and  the doped layer using their different
magnetic field dependences. What we found is that the second
contribution significantly fluctuates from one experimental run to
another. A preliminary analysis has lead us to a conclusion that
what we observe is a \emph{systematic} dependence on cooling
procedure rather than random fluctuations - the sample somehow
``remembers'' the experiment preparation conditions. Furthermore,
the doped layer contribution appears also sensitive to the sample
illumination.

The aim of the present paper is a systematic study of the ``memory
effects'' in Si $\delta$-doped and modulated doped
GaAs/Al$_{0.3}$Ga$_{0.7}$As heterostructures by the acoustic
method for different cooling and illumination procedures. We will
show that both the ac conductivity $\sigma_{xx} (\omega)$ and
sheet electron density in the interface layer,  $n_s$,  are
influenced by the above mentioned procedures. However, in a rather
wide region of parameters there exists a \emph{universal}
correlation between $\sigma_1 (\omega)$ and $n_s$ irrespectively
of the way to produce a given $n_s$. A consistent qualitative
explanation of the observed phenomena can be achieved by a
suggestion that the memory effects are due to the so-called $DX^-$
centers~\cite{6,6a} which are two-electron localized states bound by
a negative correlation energy, see, e. g., Ref.~\onlinecite{5} for
a review.

The paper is organized as follows. In Sec.~\ref{Experiment}
experimental procedure and main results are described. These results
are discussed in Sec.~\ref{Discussion}, the conclusions are drawn in Sec.~\ref{Conclusion}.

\section{Experiment} \label{Experiment}

\subsection{Experimental setup}

A SAW has been induced in a LiNbO$_3$ crystal on top of which the
sample has been placed and fixed by a spring. Details of experimental
setup are given in Ref.~\onlinecite{1}. Its important feature is that
there is no direct mechanical coupling between the LiNbO$_3$ substrate and the
sample because some finite clearance. Consequently, only electrical
coupling is present that makes determining of $\sigma_{xx} (\omega)$
simpler and more reliable.

\begin{figure}[h]
\centerline{
\includegraphics[width=\columnwidth]{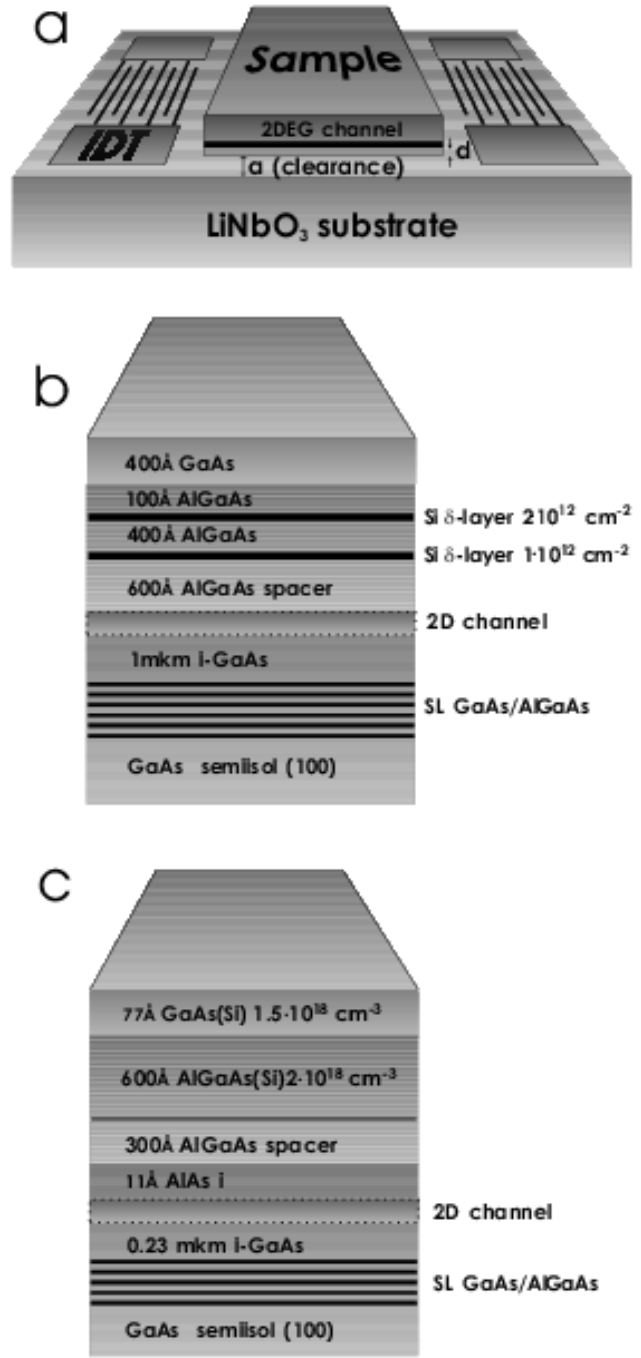}
}
 \caption{(a) Three-layers model used to extract ac conductance from acoustical data;
 Schematic structures of a Si
$\delta$-doped sample with $n_s \approx 1.5 \times 10^{11}$
cm$^{-2}$ (b), and a modulated-doped one with $n_s \approx 2.4
\times 10^{11}$ cm$^{-2}$ (c). \label{fig:01a}}
\end{figure}

Simultaneous measurements of the attenuation, $\Gamma$, and
relative variation of the sound velocity, $\Delta V (H)/V$ of SAWs
in the frequency range $f=\omega/2\pi= 30-150$ MHz were performed
in magnetic fields up to 7 T at two temperatures 4.2 and 1.5 K.
Two sorts of MBE grown GaAs/Al$_{0.3}$Ga$_{0.7}$As
heterostructures we studied: (i) Si $\delta$-doped with $n_s
\approx (1.3-4)\times 10^{11}$ cm$^{-2}$ and (ii) Si
modulated-doped heterostructures with  $n_s \approx (2.4-7)\times
10^{11}$ cm$^{-2}$. The electron densities were determined from
the periods of Shubnikov-de Haas-type oscillations of $\Gamma$ and
$\Delta V/V$ at 4.2 and 1.5 K. The structure of the samples is
schematically shown in Fig.~\ref{fig:01a}

The heterostructure was considered as a multilayer system
containing a thin conductive layer with complex sheet conductance
$\sigma_{xx} (\omega)$ at some distance $d$ from the bottom plane.
Since the distance between the interface and the doped layer is
much less than the SAW penetration depth $k^{-1}$ the quantity
$\sigma_{xx} (\omega)$ is actually the \emph{effective
conductance} of the region including interface and doped layers,
connected ``in parallel''. Dielectric constants of two other
layers (GaAs and  Al$_{0.3}$Ga$_{0.7}$As, respectively) are
assumed to be equal and denoted as $\varepsilon_s=12$. The
heterostructure is placed on top of the LiNbO$_3$ platelet with
effective dielectric constant $\varepsilon_p=50$, the vacuum
clearance between the heterostructure and the plate being denoted
as $a$. The clearance remains finite despite of the fact that the
heterostructure was pressed to the piezoelectric platelet because
of some roughness of both surfaces. Since the actual clearance is
hardly controlled, the quantity $a$ is treated as an adjustable
parameter. It is determined by fitting the experimental data at
the steps between the Hall plateaus where the conductance is
metallic and essentially frequency independent in the frequency
region of our experiments. The values of $a$ are slightly
different for different sample setups. For our experimental setup,
$a \approx (1-5) \times 10^{-5}$ cm.

According to the above model, components of the complex
conductivity have been extracted from the experimental data from
the expressions~\cite{4}
\begin{eqnarray}
\Gamma&=&4.34k
A(k)K^2\frac{\Sigma_1}{(1+\Sigma_2)^2+\Sigma_1^2}\, , \ \text{dB/cm}\,
, \label{gamma}  \\
\frac{\Delta V}{V}&=&
0.5A(k)K^2\frac{1+\Sigma_2}{(1+\Sigma_2)^2+\Sigma_1^2}\, ,
\label{dv} \\
&&\Sigma_{1,2} =4\pi t(k) \sigma_{1.2}(\omega)/\varepsilon_s V\, .
\label{sigma}
\end{eqnarray}
Here $K$ is the electromechanical coupling constant for LiNbO$_3$
(Y-cut), $A(k)$ and $t(k)$ are dimensionless functions allowing
for electrical and geometrical properties of the sample,~\cite{4}
\begin{eqnarray*}
A(k)&=& 
8 (\varepsilon_p+1)\varepsilon_s
e^{-2 k(a+d)}/\left( b_1(k)[b_2(k)-b_3(k)]\right)
  \, ,  \\
t(k)&=&[b_2(k)-b_3(k)]/2b_1(k)\, , \\
b_1(k)&=&(\varepsilon_p+1)(\varepsilon_s+1)
- (\varepsilon_p-1)
(\varepsilon_s-1)e^{-2ka}\, ,  \\
b_2(k)&=&(\varepsilon_p+1)(\varepsilon_s+1)
- (\varepsilon_p+1)
(\varepsilon_s-1)e^{-2kd}\, ,  \\
b_3(k)&=&e^{-2ka}\! \! \left[(\varepsilon_p-1)(\varepsilon_s-1)
+(\varepsilon_p-1)
(\varepsilon_s+1)e^{-2kd}\right] .
\end{eqnarray*}

\subsection{Dependence of cooling rate}

The acoustic measurements require the sample to be placed either
in vacuum, or in a dilute gas. Otherwise SAW are strongly damped
by the cooling liquid. In our experiment, the system consisting of
the sample mounted on the LiNbO$_3$ plate was placed on a cooling
finger located in a chamber. The chamber was, in turn, placed in a
He$^4$ cryostat which can be pumped out to decrease its
temperature. The superconductor solenoid in the cryostat was
cooled by liquid nitrogen.  To reach the temperatures $1.5 - 4.2$
K a dilute exchange gas (He$^4$, $\sim$ 0.1 Torr) was inserted
into the chamber, and that was the way to control the cooling
rate.

The cooling procedure was as follows. The chamber containing the
experimental setup has been initially cooled inside the cryostat
by cold gaseous He$^4$, and then liquid He has been poured into the cryostat.

Different cooling regimes were studied, and in the following they
will be referred to as \emph{slow} and \emph{rapid} cooling. In
the first case the exchange gas has been inserted from the very
beginning, at the room temperature. Then the chamber was cooled
during 1.5 - 2 hours by cold gaseous He$^4$ down to 7 - 8 K, and
finally liquid He was poured. In the second case, the chamber was
initially evacuated and cooled first by gaseous and then by liquid
He$^4$, the sample temperature being monitored by a carbon
thermometer. At some temperature $T_0$, which actually depends on
the pressure in the chamber, the exchange gas was inserted, and
the sample cooled down to the cryostat temperature during 5-10
minutes. As a result, the cooling rate is an increasing function
of $T_0$, the maximum cooling rate being at $T_0 \approx 77$ K.

\begin{figure}[h]
\centerline{
\includegraphics[width=\columnwidth]{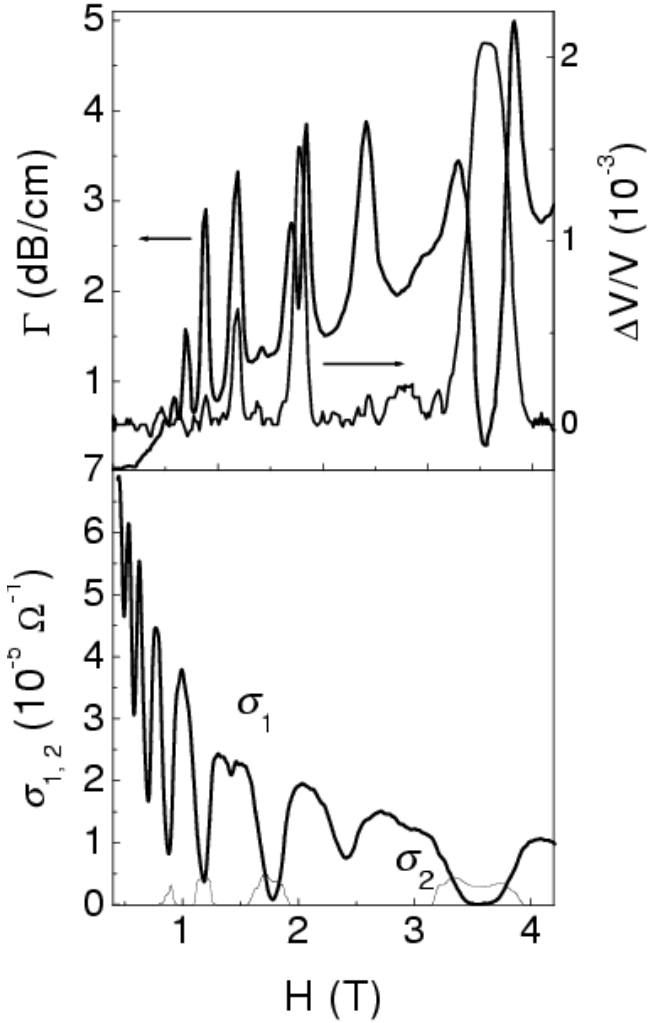}
} \caption{Top panel: magnetic field dependences of the SAW
attenuation, $\Gamma$, and of the relative velocity change,
$\Delta V/V$ for $f=30$ MHz. Bottom panel: components $\sigma_1$
and $\sigma_2$ of ac conductivity versus  magnetic field $H$ at
$T=1.5$ K. Sample --  Si $\delta$-doped
GaAs/Al$_{0.3}$Ga$_{0.7}$As heterostructure, initial electron
density $n_s \simeq 1.5 \times 10^{11}$ cm$^{-2}$. \label{fig:02}
}
\end{figure}

The results for a slowly-cooled of the Si $\delta$-doped sample
with the initial sheet electron density $n_s \simeq 1.5 \times
10^{11}$ cm$^{-2}$ are shown in the bottom panel of
Fig.~\ref{fig:02}. They are extracted from the raw data shown in
the top panel.

\begin{figure}[h]
\centerline{
\includegraphics[width=\columnwidth]{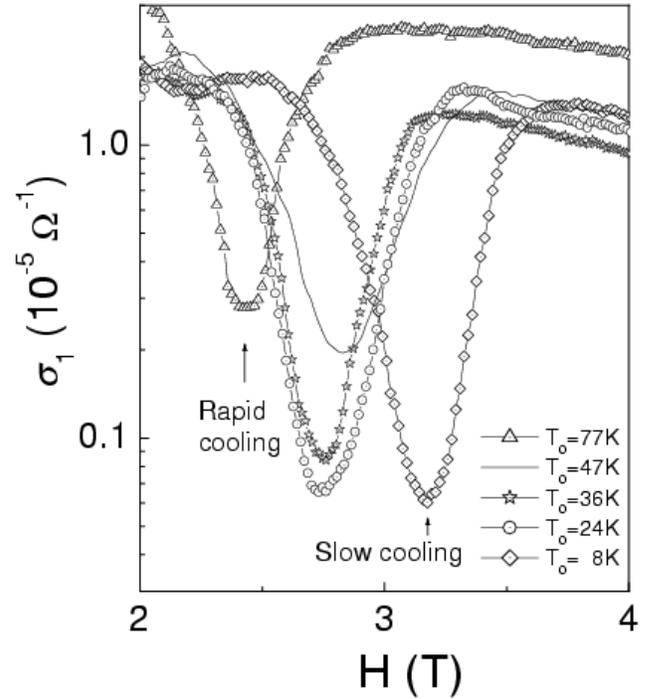}
} \caption{ Magnetic field dependences of $\sigma_1$ for $T=1.5$
K, $H=2-4$ T and different pre-cooling temperatures, $T_0$. All
curves correspond to the filling factor $\nu =2$. \label{fig:03}}
\end{figure}

Magnetic field dependences for different cooling rates extracted
at the same acoustic frequency, $f=30$ MHz, are shown in
Fig.~\ref{fig:03}. One can observe that, depending on $T_0$, (i)
minimum values of $\sigma_1$ are different, and (ii) minima occur
at different values of magnetic field. Since all the minima
correspond to the filling factor $\nu =2$, it follows that the
sheet electron density, $n_s$, is a function of $T_0$. The
behavior of $\sigma_1$ for $\nu =1$ and $\nu =4$ is similar.

\begin{figure}[h]
\centerline{
\includegraphics[width=\columnwidth]{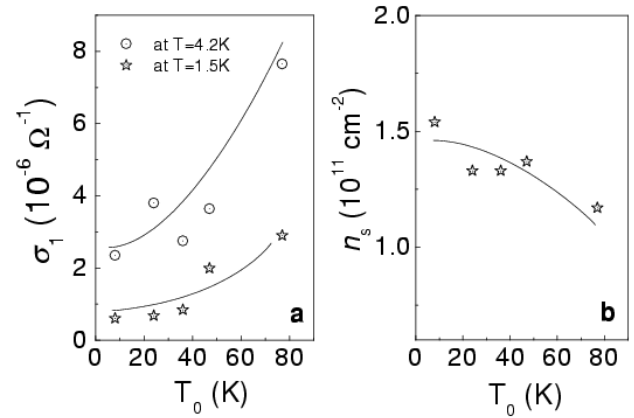}
} \vspace*{-2mm} \caption{Dependences  of the ac conductance
$\sigma_1$ (left panel) and sheet electron density $n_s$ (right
panel) on the pre-cooling temperature $T_0$. $\nu=2$, temperatures
of experiment are shown in the legend. \label{fig:04}}
\end{figure}

In Fig.~\ref{fig:04}, the values of $\sigma_1|_{\nu=2}$ and $n_s
|_{\nu=2}$ are plotted versus the pre-cooling temperature, $T_0$.
The first quantity increases with increase of $T_0$, while the
second one decreases. For all $T_0$, imaginary part of the
conductivity, $\sigma_2$, remains greater that $\sigma_1$ that
indicates hopping conductance. $\sigma_2$ is also dependent on
$T_0$, however its dependence is much slower than that of
$\sigma_1$. The general behavior of the ac conductance in the
modulated-doped samples is similar.

The dependence of the sheet electron density $n_s$ turns out to be
correlated with the \emph{initial} value of $n_s$; the larger the
initial $n_s$ the slower its dependence on $T_0$. For the sample
with $n_s \simeq 1.5 \times10^{11}$ cm$^{-2}$ the relative
variation of $n_s$ is 30\%, for $n_s \simeq 2.3 \times10^{11}$
cm$^{-2}$ is 10\%, while $n_s=7 \times10^{11}$ cm$^{-2}$ the
dependence on $T_0$ is essentially absent. In the last sample the
first band of size quantization is almost full.

It is worth noting that the state reached by fast cooling does not
change for a long time. In particular, characteristics of the
sample with $n_s=1.2 \times10^{11}$ cm$^{-2}$ after maximally fast
cooling from $T_0=77$ K did not change during at least 28 hours.

\subsection{Dependence on illumination}

\begin{figure}[h]
\centerline{
\includegraphics[width=\columnwidth]{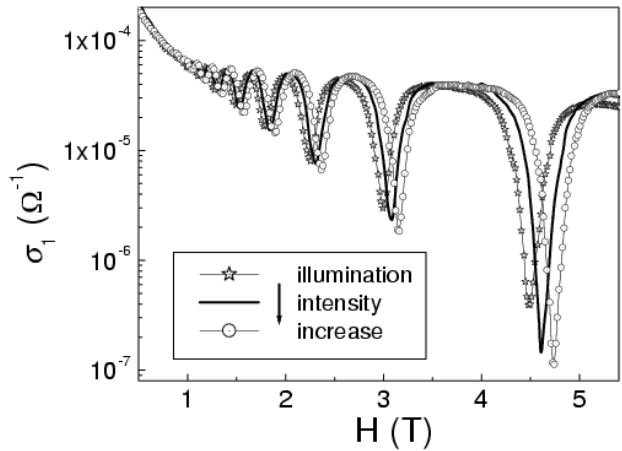}
} \vspace*{-2mm} \caption{Magnetic field dependences of $\sigma_1
(\omega)$ after successive  illumination, $\lambda=0.81$~$\mu$m,
$T=4.2$~K, $f=30$~MHz.
 \label{fig:05}}
\end{figure}

\begin{figure}[h]
\vspace*{2mm}
\centerline{
\includegraphics[width=\columnwidth]{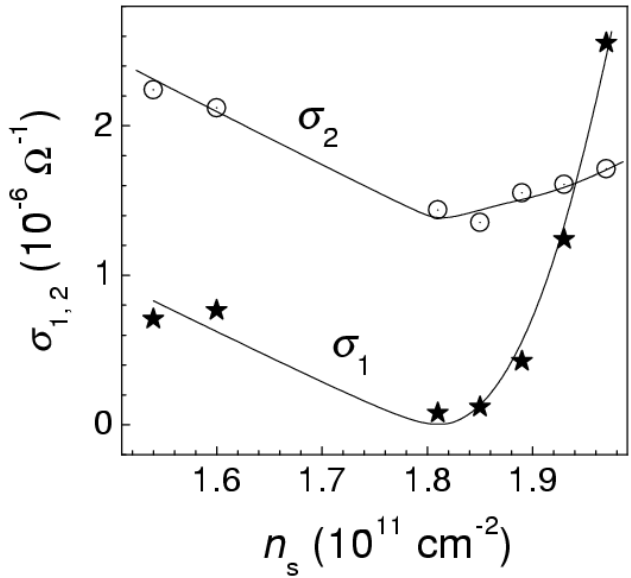}
}
\vspace*{-2mm}
\caption{Effect of illumination presented as $\sigma_{1,2}$ versus
  $n_s$ curves. Each point corresponds to a state after a given illumination
  dose, $\nu=2$.  \label{fig:06}}
\end{figure}

As well known, dc conductance of Al$_{0.3}$Ga$_{0.7}$As thin film
~\cite{6} and GaAs/Al$_{0.3}$Ga$_{0.7}$As heterostructures
~\cite{Buks} is sensitive to infrared (IR) illumination.
 To investigate combined effect of slow cooling and IR illumination on ac conductance
  light emitting diodes (LEDs) producing IR radiation with the wavelength
$\lambda = 0.81, 1.34, 2.53$ and $4.47$~$\mu$m were placed into
the chamber \cite{Matveev}. Choosing the illumination dose we were
able to change the sheet electron density, as well as the ac
conductance, by controllable portions.
 The persistent ac photoconductance is
observed only if the illumination frequency exceeds some threshold
located between 0.92 and 0.49 eV ($\lambda =1.34$ and
$2.53$~$\mu$m, respectively). The magnetic field dependence of
$\sigma_1 (\omega)$ after successive illumination of a
$\delta$-doped sample with $n_s \simeq 3.3 \times
10^{11}$~cm$^{-2}$ is shown in Fig.~\ref{fig:05}. One can see that
successive pulsed illumination (pulse duration $\lesssim 10$~s)
leads to a shift in the location of the corresponding $\sigma_1$
minimum. The shift indicates an \emph{increase} in the 2DEG
density, $n_s$. At the same time the minimum value of $\sigma_1$
\emph{decreases}.
The effect of the illumination in a Si doped structure
 is illustrated in
Fig.~\ref{fig:06}. Here $\sigma_1 (\omega)$ and  $\sigma_2
(\omega)$ are plotted as functions of the sheet electron density
$n_s$, tuned by successive illumination. All three quantities are
extracted from acoustic measurements at $T=1.5$~K. One can see
that increase in $n_s$ is accompanied by decrease of $\sigma_1$
down to to some value after which a rapid increase occurs. In this
regime the successive illumination leads to an increase in
$\sigma_1$ leaving $n_s$ almost constant.

\section{Discussion} \label{Discussion}

\begin{figure}[h]
\centerline{
\includegraphics[width=\columnwidth,angle=0]{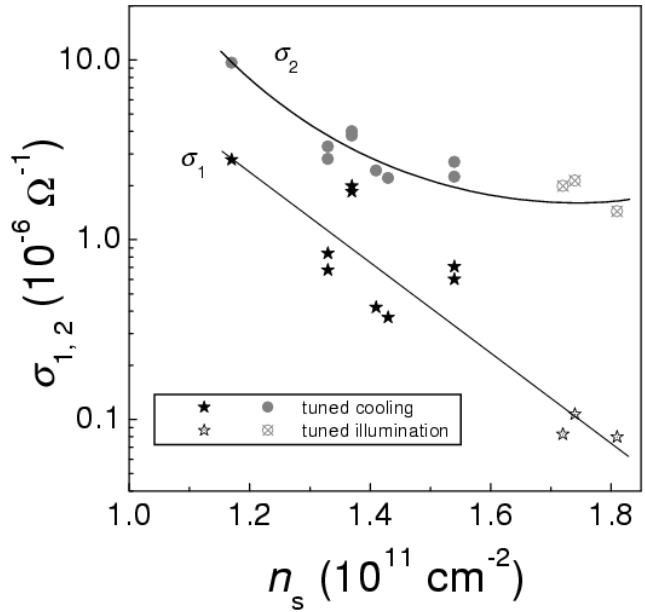}
}
\caption{Combined effect of cooling and successive illumination for
  $\sigma_1(\omega)$ and  $\sigma_2(\omega)$ which are plotted versus
  $n_s$ tuned either by fast cooling, or by illumination. $\delta$-doped sample,
  $n_s \simeq 1.5 \times 10^{11}$~cm$^{-2}$,  $T=1.5$~K. Measurements are performed at
  $f=30$~MHz; $\nu =2$. \label{fig:07}}
\end{figure}

As we expected,~\cite{4} the set of experimental results indicates
a significant role of the doped layer in the ``memory" effects.
Indeed, a remarkable feature of the experimental results is that,
being presented as $\sigma_1$ and $\sigma_2$ versus $n_s$, they
are more-or-less universal. Namely, the data obtained for
different cooling and illumination procedures \emph{collapse} to
almost same curves. To demonstrate this feature, in
Fig.~\ref{fig:07} we plot $\sigma_1$ and $\sigma_2$ versus $n_s$
for a Si  $\delta$-doped sample with initial electron
concentration $n_s \simeq 1.5 \times 10^{11}$~cm$^{-2}$. Here the
results both for different pre-cool temperatures $T_0$ and
different illumination doses are included.
We believe that such property, as well as the threshold in ac
photoconductance support the idea that the electron states in the
doped layer are just the so-called $DX^-$ centers.~\cite{6} These
states are actually two-electron bound states stabilized by local
lattice distortion.  $DX^-$centers were observed both in
GaAs/Al$_{0.3}$Ga$_{0.7}$As heterostructures and in
Al$_x$Ga$_{1-x}$As films with $x > 0.22$. They are considered  to
be responsible for dc persistent photoconductance in these
systems, which has also a threshold in the illumination quanta
energy located between 0.6\cite{6} and 0.8~eV \cite{7}.

In general, the defects responsible for the $DX^-$-centers have
three charge states which differ by number of electrons occupying
the center. Due to a local lattice distortion the two-electron
state has the lowest energy if the two electron correlation energy
$|U|$ exceeds the thermal energy $kT$.\cite{comment} As a result,
at $T \ll |U|/k$ the the defects are either occupied by two
electrons and negatively charged ($D_-$-centers)  or empty and
positively charged ($D_+$-centers) . The $D_-$-centers can be
treated as \emph{small bipolarons}. The real situation is more
complicated than the one discussed above since there is no unique
opinion on the microscopic nature of excited states of the defect. In
particular, there are several intermediate states of the defect each
containing one electron, but differing by the amount of lattice
distortion. In the following we will not discriminate between these
states since detailed theory of acoustically-stimulated electron
tunneling is beyond the scope of the present work. To simplify the
discussion, following Ref.~\onlinecite{8},  we assume that a set of
equal number of $D_-$ 
and $D_+$ states form the ground state which is the only state
occupied at very low temperature. The one-electron (neutral)
state $D_0$ has the energy higher than the ground state energy by
the correlation energy o$|U|$.

In heterostructures, one can imagine several processes taking
place during a cooling from the room temperature to 1.5~K. Among
them are freeze-out of the carriers in the Si-doped layer of
Al$_{0.3}$Ga$_{0.7}$As to occupy first deep $D_-$ states and then
shallow states,\cite{9} carriers exchange between the doped and
the interface layers, etc. We believe that under rapid cooling the
electron distribution at the pre-cooling temperature $T_0$ is
frozen - it does not significantly change during the cooling below
$T_0$. This statement is compatible with a crude estimate of the
activation energy based on the expression $E_d \approx 0.7x-0.15$
eV which has been obtained for Al$_x$Ga$_{1-x}$As films for $0.22
<x < 0.4$.~\cite{9} For our samples, this estimate yields $E_d =
60$~meV. Consequently, $e^{-E_d/k T}$ varies from $10^{-1}$ to
$10^{-4}$ as $T$ varies from 300 to 100 K.

Acoustic methods under conditions of the QHE provide a unique
possibility to separate the contributions of the interface and
doped layers. Indeed, at the QHE plateaus the electronic states in
the interface layer are \emph{localized}, and their contribution
to $\sigma_{xx}$ are small. As a result, the contribution of the
electron hopping in the \emph{doped layer}  becomes
measurable.~\cite{4} We believe that the main mechanism leading to
this contribution is due to tunneling transition of electron pairs
(bipolarons) between a $D_-$ center and an adjacent $D_+$ one. At
low temperatures sequential tunneling via the neutral states $D_0$
is probably not important since the energy difference between
$D_0$ and the ground state is of the order of $|U| \approx 1$~eV.
However, $D_0$ states play an important role in the formation of
the tunneling barrier between $D_-$ and $D_+$ centers. A theory of
ac hopping conductance due to $DX^-$-centers in three-dimensional
amorphous materials has been developed in Ref.~\onlinecite{8}.
According to this theory, is is rather difficult to discriminate
between single-electron and two-electron tunneling from frequency
and temperature dependences of $\sigma (\omega)$. We are not aware
of a theory of ac hopping conductance relevant to pair tunneling
in GaAs/Al$_{0.3}$Ga$_{0.7}$As heterostructures. However, one can
expect that the differences in frequency and temperature
dependences are very small also in this case.

Assuming that the doped layer's thickness is much less than $k^{-1}$
we can use for $\sigma_1$ an estimate similar to the well-known
expression for the single-electron tunneling,~\cite{3}
$$\sigma_1(\omega) \sim g_b^2 \xi^3_b \omega e^4/\varepsilon_s\, .$$
Here $g_b$ is the density of states for bipolarons, $\xi_b$ is
their localization length, while $e$ is the electron charge. We
believe that it is the density of $D_-$ states, $g_b$, that is
influenced by cooling and illumination procedures. In particular,
the IR illumination causes ionization of the $D_-$ centers in the
doped layer. Part of released electrons tunnels into the interface
layer that results in the increase in the sheet density $n_s$. A
similar effect occurs during a rapid cooling due to quenching of
the electron transfer between the defects and 2D layer. The
higher the pre-cooling temperature $T_0$ the less time left for
the electrons to tunnel into the interface layer. As a result,
$n_s$ is a decreasing function of $T_0$.

It worth mentioning that there is an important difference between
the dc persistent photoconductance and the effect we observe.
Indeed, in the first case the conductance is possibly due to
extended electron states either in the 2DEG-layer or in the
conduction band in Al$_{0.3}$Ga$_{0.7}$As layer. Contrary, the
states responsible for the ac conductance are \emph{localized}
that follows from the relation $\sigma_2 (\omega) \gtrsim \sigma_1
(\omega)$. Illumination leads to a \emph{decrease} in the number
of occupied bound states. As a result, the ac conductance
decreases with illumination. At very high illumination dose the
$DX^-$ centers become ionized, and all the ac conductance is due
to extended states.

We have employed the procedure of Ref.~\onlinecite{4} to separate
the contributions of the doped layer and of the 2D layer in the
ac hopping conductance and then analyzed the latter using a
picture of single electron nearest-neighbor tunneling. Tuning the
2DEG density by illumination after slow cooling to change the
magnetic field $H$ corresponding $\nu =2$ from 2.7 to 3.8~T we
have found that the localization length $\xi$ in this region
behaves as $H^{-1/2}$. This dependence is compatible with the
assumption of the single electron nearest neighbor tunneling.
Unfortunately, we were not able to carry out a similar analysis
for $\nu=4$ since at large $\nu$ the components $\sigma_1$ and
$\sigma_2$ are of the same order of magnitude, and the mechanism
of ac conductance is mixed.

\bigskip
\section{Conclusions} \label{Conclusion}

 Main conclusions from the present work can be formulated as follows.
\begin{itemize}
\item [(i)] Both ac hopping conductance and sheet electron density in Si
 $\delta$-doped and  modulation doped  GaAs/Al$_{0.3}$Ga$_{0.7}$As
 heterostructures at the quantum Hall effect plateaus depend on the
 samples' \emph{cooling rate}.

\item [(ii)] Successive IR illumination leads to a \emph{persistent ac hopping
    photoconductance} which decreases with the illumination dose. At
  the same time, the sheet electron concentration, $n_s$, in the
  interface layer increases. The persistent ac photoconductance occurs only if the illumination
  frequency exceeds some threshold located between 0.5 and
  0.9~eV.

\item [(iii)] The above set of results can be qualitatively
  interpreted within the framework of the concept of $DX^-$ centers --
  localized two-electron states bounded by local lattice
  distortion -- located in the doped layer of the heterostructure. We
  believe that the doped layer ac hopping conductance is due to
  tunneling of electron pairs (bipolarons) between adjacent doubly occupied
  and empty defects.

\item [(iv)] The contribution of tunneling between localized states in
  the interface layer can be extracted and used to estimate  the
  electron localization length, $\xi$. Within the experimentally-accessible
  range, $\xi \propto H^{-1/2}$ which is compatible with
  single-electron nearest neighbor tunneling.
\end{itemize}

\acknowledgments One of the authors (ILD) is thankful B.~A.~Volkov
and D.~R.~Khokhlov for discussions. The work is supported by RFFI
01-02-17891, MinNauki, Presidium RAN grants and Russia-Ukraine
Program {\em Nanophysics and Nanoelectronics}.


\end{document}